\documentclass{etc/class}

\usepackage{amsmath, amsthm, amssymb, amsfonts} 
\usepackage{mathtools}                          
\usepackage{bbm}                                
\usepackage{tkz-tab}                            
\usepackage[caption=false]{subfig}
\usepackage{cancel}                             
\usepackage{algorithm}                      
\usepackage{algpseudocode}                  

\usepackage{physics}                            

\usepackage{algorithm}                          
\usepackage{algpseudocode}                      

\usepackage{multirow}                           

\usepackage{tikz}                               
\usetikzlibrary{tikzmark,calc}
\usepackage{wrapfig}                            
\usepackage{svg}                                
\usepackage{float}

\usepackage{lipsum}                             
\usepackage{xcolor}                             
\usepackage{soul}                               
\usepackage[super]{nth}                         
\usepackage{footnote}
\usepackage{ulem}

\usepackage{xparse}                             
\usepackage[capitalize]{cleveref}

\newcommand{\id}{\mathbbm{1}}

\newcommand{\fid}{\mathcal{F}}

\newcommand\steparrow[5][0.1]%
  {\begin{tikzpicture}[remember picture,overlay]
   \draw[-stealth]
     ($({pic cs:#4}|-{pic cs:#2})+(#1,0)$)
     .. controls +(0.2,-0.05) and +(0.2,0.1) ..
     node[right,align=center]{#5}
     ($({pic cs:#4}|-{pic cs:#3})+(#1,0.1)$);
   \end{tikzpicture}%
  }

\begin{document}


\title{
	Unitary Synthesis with AlphaZero via Dynamic Circuits
}

\author{Xavier Valcarce}
\email{xavier.valcarce@ipht.fr}
\affiliation{Universit\'e Paris-Saclay, CEA, CNRS, Institut de physique th\'eorique, 91191, Gif-sur-Yvette, France}
\author{Bastien Grivet}
\affiliation{Universit\'e Paris-Saclay, CEA, CNRS, Institut de physique th\'eorique, 91191, Gif-sur-Yvette, France}
\author{Nicolas Sangouard}
\affiliation{Universit\'e Paris-Saclay, CEA, CNRS, Institut de physique th\'eorique, 91191, Gif-sur-Yvette, France}

\begin{abstract}
	Unitary synthesis is the process of decomposing a target unitary transformation into a sequence of quantum gates. This is a challenging task, as the number of possible gate combinations grows exponentially with the circuit depth.
	In this manuscript, we propose an approach using an AlphaZero-inspired reinforcement-learning agent for the exact compilation of unitaries using discrete sets of logic gates.
	The approach achieves low inference time and proves versatile across different gate sets, and qubit connectivities. Leveraging this flexibility, we explore unitary synthesis with dynamic circuits -- circuits that contain non-unitary operations such as measurements and conditional gates --
	and discover unusual implementations of logical quantum gates. Although the direct synthesis of complete algorithms is intractable, our approach is well suited for efficiently synthesizing subroutines. This may have a significant impact when these subroutines are invoked repeatedly during algorithm execution.
\end{abstract}

\maketitle


\section{Introduction}

Quantum computing systems must integrate diverse software and hardware components to bridge quantum applications and quantum processors. Compilers in particular, play a key role in the quantum computing full-stack to translate high-level programming languages defining logical sequences into assembly-like instructions executable by the processor~\cite{Maronese2022}. A critical aspect of quantum compilation is unitary synthesis~\cite{Dawson2006,Smith2023,Younis2021,Paradis2024}, which aims to design a sequence of executable elementary operations such as logical quantum gates that matches or approximates a target unitary matrix.

The target unitary typically corresponds to the evolution of a small algorithmic subroutine rather than that of an entire algorithm. Indeed, executing quantum algorithms for practical problems generally requires between $10^7$ and $10^{11}$ logical quantum gates~\cite{Beverland2022, Daley2022, Hoefler2023, Dalzell2025}. Since the number of possible gate combinations grows exponentially with circuit depth, brute-force synthesis of complete algorithms is infeasible. A more practical strategy is therefore to synthesize smaller subroutines that can be reused repeatedly during algorithm execution. For instance, Shor’s algorithm for factoring $n$-bit RSA integers requires $o(n^2)$ additions of $n$-bit numbers~\cite{Shor1994}, with each addition typically decomposed into $o(n)$ two-bit additions~\cite{Gidney2021,Gouzien2021}. Consequently, an appropriate synthesis of the two-bit addition primitive -- or of its constituent quantum gates -- can substantially reduce both the runtime and qubit overhead of the entire factorization algorithm.

If the goal is not to design a unitary synthesizer capable of handling extremely large unitaries, it must instead be versatile and efficient. A versatile synthesizer produces quantum circuits that adhere to the constraints of the physical computing architecture. For instance, the choice of quantum gates and qubit connectivity must align with the physical platform and processor layout. An efficient unitary synthesizer must also include mechanisms to reduce both the runtime and qubit overhead of quantum subroutines. Interesting techniques include dynamic circuits -- circuits that contain non-unitary operations such as measurements on ancillary qubits and conditional gates.
They have been shown to lower the number of non-Clifford gates~\cite{Selinger2013,Jones2013,Gidney2018}, which are substantially more costly to implement than Clifford gates and often dominate the overall execution cost of a circuit~\cite{Fowler2012, Litinski2019, Gouzien2023}.

In this manuscript, we investigate a reinforcement learning approach to exact unitary synthesis with discrete gate sets, using an AlphaZero-like agent~\cite{Silver2018}. Building on recent results~\cite{Rietsch2024} demonstrating that AlphaZero variants enables versatile unitary synthesis, we show that a model-based reinforcement learning utilizing Monte Carlo tree search achieves both versatility and efficiency.
Concretely, we train learning agents to synthesize unitaries on up to three qubits, with and without an additional ancilla, using a standard gate set consisting of Clifford and T gates. After training, we show that our method can, on average, synthesize unitaries into circuits with shorter depths than those of randomly generated circuits of up to depth 40 used to define the target unitaries. When the agent is tasked with synthesizing structured unitaries such as the Toffoli and Control-S gates, we show that it recovers within a few seconds, implementations that are known to be optimal in terms of the number of T-gates for almost all examples. The cases with the ancillae and unitary synthesis conditioned on the measurement of this ancilla lead to unusual implementations of these gates, including a $4-$T gate implementation of the Toffoli gate, all applied on the ancillae.

\medbreak

We begin by establishing the framework for exact unitary synthesis with a discrete gate set and provide a comprehensive review of available techniques, discussing their respective advantages and limitations in \cref{sec:eus}. We then present the preliminaries on reinforcement learning and the AlphaZero agent in \cref{sec:rl}. \cref{sec:eus+rl} discusses the proposed method to tackle exact unitary synthesis with an AlphaZero-like agent. We then expose results in \cref{sec:results} before concluding on the relevance AlphaZero-like agents for exact unitary synthesis.

\section{Exact Unitary Synthesis}
\label{sec:eus}

\begin{figure*}
    \centering
    \includegraphics[width=\textwidth]{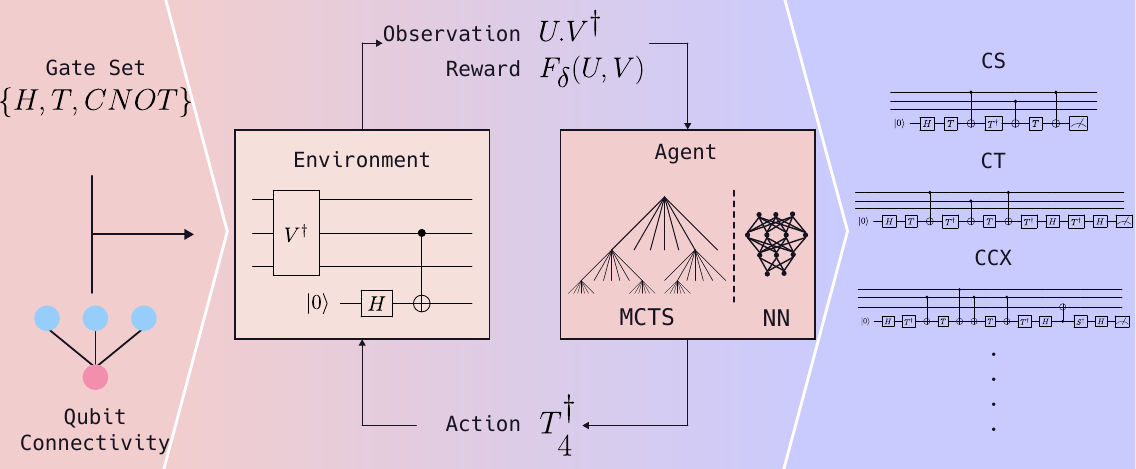}
    \caption{Exact unitary synthesis using and AlphaZero-inspired agent. A gate set and qubit connectivity are specified. The reinforcement learning (RL) agent learns to synthesize unitaries, using the specified architecture, by trial-and-error on many randomly sampled target unitaries $V$. After training, the agent can synthesize some provided target unitaries.}
    \label{fig:EUS_RL}
\end{figure*}

In unitary synthesis, the task is to implement a target operation on $n$ qubits, specified by a unitary $V\in \mathcal{C}^{2n\times 2n}$. We are given a set of quantum gates $\mathcal{G}$, which may depend on the underlying physical platform and the chosen error-correction scheme. The connectivity and physical layout then determine the subset of elementary operations $\mathcal{G}$ that are effectively available for constructing $V$. For $G=\{H,T,CNOT\}$ for example, a 1D layout of three qubits with nearest-neighbour connectivity yields $\mathcal{G}=\{\text{H}_1,\text{H}_2,\text{H}_3,\text{T}_1,\text{T}_2,\text{T}_3,\text{CNOT}_{12},\text{CNOT}_{23}\}$, where the subscripts indicate the qubits on which the gates act. Importantly, we also consider operations that make use of ancillaes, which are initialized in the $|0\rangle$
state and projected into the $|0\rangle$ state  by post-selection.
Such clean-ancilla circuits may be probabilistic, as ancillae can be measured in the $\ket{1}$ state.
When this occurs, a classically-controlled correction unitary -- applied conditionally upon detecting $\ket{1}$ -- can be used to recover the target unitary on the data qubits.
Circuits implementing these correction unitaries can be synthesized in a second step, once a circuit associated with the ancilla outcome $\ket{1}$ has been identified. The correction $U_C$ can be found by targeting the synthesis of the unitary $U_{\ket{1}}^\dag V$ where $U_{\ket{1}}$ is the unitary produced when $\ket{1}$ is measured. This allows us to consider dynamic circuits to synthesize unitaries. 

We denote by $U$ the matrix corresponding to the circuit constructed from a sequence of gates in $\mathcal{G}$, with or without ancillas and measurements. The goal of unitary synthesis is to find a circuit such that $F(U,V)=\Tr(V^\dagger U) = 1-\varepsilon$. 
Exact unitary synthesis corresponds to the case $\varepsilon=0$.
 
\medbreak

To synthesize a unitary exactly using a finite gate set, the most common approaches leverage the meet-in-the-middle search technique. 
Ref.~\cite{Amy2013} first introduced this method, synthesizing T-depth optimal circuits and supporting ancilla qubits, but suffering from prohibitively slow runtime.
Subsequent improvements focused on achieving T-count optimality and shortening runtime at the cost of dropping the support for ancillae~\cite{Gosset2013,Matteo2016}.
More recently, Refs~\cite{Mosca2021,Gheorghiu2022} have refined meet-in-the-middle routines to further optimize runtimes, targeting T-count and T-depth optimization, respectively.
However, these two recent works are not compatible with dynamic circuits, sacrifice optimality, and still exhibit relatively high runtimes.

Most recently, Synthetiq -- an algorithm based on simulated annealing -- has demonstrated the synthesis of resource-efficient circuits with runtimes orders of magnitude faster than the best meet-in-the-middle methods~\cite{Paradis2024}. This approach supports custom gate sets and clean ancilla architectures; however, it does not account for specific qubit connectivities~\cite{SynthetiqConnectivity}.

Other notable approaches include the one reported in Ref.~\cite{Kang2023} which employs enumerative search with pruning. While it supports custom gatesets, it is not compatible with the use of ancillae.
Besides, method based on SAT solvers have been proposed, see Ref.~\cite{Gouzien2025} and references therein. 
Although they support ancillae and guarantees circuit depth optimality, their runtimes may be prohibitive.

\medbreak

Machine learning has found many successful applications in quantum computing, with most visible recent results focusing on the development of high-accuracy decoders~\cite{Bausch2024}. Reinforcement learning, in particular, proves to be useful for the synthesis of single-qubit~\cite{Alam2023} or target unitaries~\cite{He2021}, the preparation of two-qubit quantum states~\cite{Kolle2024}, the synthesis of large Clifford circuits~\cite{Kremer2025} and the discovery of error correction codes and encoders~\cite{Olle2024}. The most closely related work is reported in Ref.~\cite{Rietsch2024}, which synthesizes Clifford+T circuits for up to five qubits, generated from randomized circuits of up to 60 gates, using Gumbel AlphaZero. Encouraged by this work, we present a reinforcement learning method inspired by AlphaZero to exactly compile unitaries. In particular, we focus on training agent for discovering dynamic circuits with constrained gates and qubit connectivities. An overview of our approach is depicted in \cref{fig:EUS_RL}.

\section{Reinforcement Learning}
\label{sec:rl}

\subsection{Generalities on Reinforcement Learning}

\textit{Single-agent reinforcement learning} is an area of machine learning where an \textit{agent} learns to navigate an \textit{environment} with predefined rules. 
The agent interacts with the environment through actions, altering its state. 
For each interaction, the environment provides feedback in the form of a reward, evaluating the performance of the action, and a current observation of its internal state. 
Through repeated interactions, the agent learns strategies to maximize the overall received reward.

Formally, an agent interacts with an environment over a \textit{game} consisting of $t$ steps.
For every step, the agent selects an action $a_t$ from an allowed action set $\mathcal{A}_s$, following a policy $\pi(a_t|s_t)$, where $s_t$ is the current state of the environment or \textit{game state}.
After executing this action, the environment returns the updated state $s_{t+1}$ and a reward $r_{t+1}$.
Over many games, the policy $\pi$ is updated, or \textit{trained}, to encourage actions that leads to the highest cumulative reward $R=\sum_t r_t$.

\subsection{AlphaZero}

Introduced by DeepMind, AlphaZero agents represent a significant advancement in model-based reinforcement learning~\cite{Silver2018}.
These agents use a hybrid version of the Monte Carlo Tree Search (MCTS), where a deep neural network is coupled with the tree search for reward and policy prediction.
This method has seen numerous successful applications in complex board games such as Chess, Go, and others~\cite{Segler2017,Moy2019,Dalgaard2020}.

The Monte Carlo Tree Search algorithm~\cite{Coulom2007} iteratively builds a decision tree, where each node represents a game state and edges represent possible actions. 
Based on a balance between exploration and exploitation, the algorithm selects well-chosen actions, aiming to explore valuable regions of the tree.

In the context of reinforcement learning, MCTS can be used as the \textit{policy provider} of an agent.
At every step $t$ of the game, the agent has access to a \textit{virtual environment} where a number $N_{MCTS}$ of roll-outs of the tree are performed, starting from the current state $s_t$ and exploring the tree and the potential outcomes of the game.
Each of the $N_{MCTS}$ roll-outs is a simulation of the game, also called \textit{virtual game}, leading to potential rewards.
Finally, the statistics from the built decision tree are used to construct a policy $\pi$. 
In \cref{app:MCTS}, we provide a detailed explanation of this algorithm.

\medbreak

While effective, MCTS faces limitations when the simulations budget is small, rewards are sparse, or the action space is large.
Moreover, MCTS is a highly computationally intensive algorithm, as a large number of virtual games are required for each selected action.
AlphaZero addresses this by hybridizing a \textit{neural network} with the MCTS.
The neural network guides the MCTS using a prior policy $\pi_{NN}$, which helps in selecting actions leading to higher rewards.
It also helps building the search tree by providing an estimated reward $V_{NN}(s_k)$ at each node $s_k$.
Overall, the neural network circumvents sparse-reward and large action space limitations, allowing to save resources by keeping a low simulation budgets, and while improving the overall performance of the constructed policy.

\medbreak

AlphaZero requires a training phase, in which the agent self-plays games using the hybridized version of the MCTS with a low simulation budget. 
Every $n_\text{games}$ played, the weights of the neural network are updated to minimize a cost function designed to improve the perceived cumulative reward.
After the resource-intensive training, AlphaZero agents can be used on specific games with an increased simulation budget to enhance their performance.
In \cref{app:Alphazero}, we provide a detailed description of the algorithm and its training routine.

\section{Reinforcement Learning for Exact Unitary Synthesis}
\label{sec:eus+rl}

\subsection{Game framework}

Exact Unitary Synthesis can be viewed as a single-player game where an agent aims to construct a quantum circuit that matches an input target unitary $V$.
Starting from an empty circuit, the agent incrementally appends gates from a fixed gate set.
Each gate added to the circuit results in a specific reward that quantifies the proximity of the circuit to the target unitary. 
Finally, the game concludes when the agent has either synthesized the target unitary or reached a given maximum depth.

\medbreak

Formally, an agent is trained for a fixed number of qubits, gate set, and qubit connectivity, which directly defines the action set $A_s$ corresponding to the gate set $\mathcal{G}$. Each action $a\in A_s$ thus represents a gate applied to specific qubits.
A game starts with the environment initialized to an empty quantum circuit and an input target unitary $V$.
Equivalently, the environment can be viewed as being initialized with a circuit with a given transversal operation $V^\dagger$ on all qubits, as depicted in \cref{fig:EUS_RL}.

At every step $t$, an action $a_t\in A_s^t$ is selected by the agent.
To avoid trivial actions, such as choosing a gate that would cancel a previous action, the action set available at step $t$, $A_s^t$, is masked to only allow non-trivial gates.
For every gate in $A_s$, we push the gate through the current constructed circuit via commutation relations.
When two gates do not commute, we check whether they form a redundancy, i.e. if they lead to the identity operation, in which case we discard the corresponding action from $A_s^t$.

The environment performs action $a_t$, simply appending the corresponding gate to the quantum circuit.
The game state returned by the environment is the matrix product $s_t=U_t.V^\dagger$, where $U_t$ is the unitary associated to the circuit obtained at step $t$.
With this framework, one game step only requires a single matrix multiplication as $s_{t+1}=U_{t+1}.V^\dagger=a_{t+1}.s_t$.
Moreover, a quantum circuit synthesizing the input target unitary leads to $s=\id$, up to a global phase, for any provided target.

The reward returned by the environment is designed to capture the closeness between $U_t$ and $V$.  An effective metric for quantifying how close two unitaries are is the fidelity~\cite{Nielsen2012}. To avoid misleading behavior, we use a binary fidelity
\begin{equation}
    \fid_\delta \left(  U, V \right) = 
    \begin{cases}
        0 \quad \text{if} ~~ &1-\Tr \left[U \cdot V^\dagger \right] > \varepsilon \\
        1 \quad \text{if} ~~ &1-\Tr \left[U \cdot V^\dagger \right] \leq \varepsilon
    \end{cases},
\end{equation}
where we set $\varepsilon=10^{-3}$ to account for numerical errors. This binary reward function ensures that the agent receives a clear signal when reaching the target unitary $V$, thereby guiding the learning process effectively.

\subsection{Reinforcement Learning Pipeline}

The reinforcement learning pipeline consists of training the agent through self-play, using different target unitaries for every game. 
The target unitaries are generated from random circuits of depth $d$, the \textit{target depth}.
These circuits are built iteratively using the same gate set available to the agent, with each gate stochastically selected following a uniform distribution.
To prevent the generation of trivial circuits that could hinder the learning process, gate redundancies are avoided using the same action mask detailed in the previous section.
This framework ensures the existence of a solution, or reward 1, in a maximum of $d$ steps.
In \cref{app:random_circuit}, we detail the construction of these circuits.

Learning from random circuit, even if a solution exists, can nevertheless be challenging; rewards are sparse and the space of possible circuits grows exponentially with the circuit depth. 
We use curriculum learning to tackle this issue, i.e. we progressively increase the \textit{difficulty} of the game as learning advances.
Concretely, we sample the target depth of circuits from a Gaussian distribution with mean $\mu$ and variance $\sigma=5$.
Every $n_\text{games}$ games, we increment $\mu$ by 1.
Furthermore, we set the maximum number of steps allowed per game to $d$, as we know at least one solution exists for that depth.
Together, this allows to save drastic resources during the learning process.
Indeed, training from smaller depth circuits first increases the number of won games to learn from, as the MCTS alone may find solutions.
Moreover, as the average target depth increases, the policy is sufficiently updated to successfully synthesize some non-trivial target unitaries, avoiding sparse rewards.
Using a step limit function of $d$ also saves computational resources while favoring agents that use shorter depth circuit to synthesize unitaries.

To further improve the learning process, we use a scheduled policy temperature, to encourage exploration in the first steps of games and exploitation for the last steps.
A higher temperature makes the action selection more stochastic, following the probability distribution of the policy $\pi(a|s)$ for $T=1$, while a lower temperature makes it more deterministic, selecting deterministically the action with the highest probability for $T=0$. 
Formally, we compute the probability of selecting an action $a$ following
\begin{equation}
    \Pi(a|s) = \begin{cases}
        \frac{\pi(a|s)^{(1/T)}}{\sum_i \pi(i|s)^{(1/T)}}, & \forall T \in (0,1] \\
        \text{argmax}_a \pi(a|s), & T=0.
    \end{cases}
\end{equation}
In practice, we set a temperature of $T=1$ for the first $5$ steps before decreasing it to $T=0.6$ for the next $10$ actions and down to $T=0.4$ for the remaining ones. 

\subsection{Implementation and Parameters}

We implement the game environment fully in JAX~\cite{jax2018}, enabling it to run on GPU/TPU accelerators, and use the PGX game framework as many machine learning libraries support this format.
Our implementation is available on GitHub\footnote{\url{https://github.com/xvalcarce/quantum_compilation}}.

On the agent side, we use a vectorized implementation of AlphaZero written in JAX, adapted from \textsc{turbozero}~\cite{turbozero}\footnote{\url{https://github.com/xvalcarce/turbozero\#qc}}.
For training purpose, we parameterize our AlphaZero agent to use MCTS with a simulation budget of $N_{MCTS}=200$ per action, balancing performance while being light on resources.
For the neural network, given the complexity of the unitary synthesis task, we opt for a two-head deep residual network architecture, or ResNet~\cite{He2015}.
ResNets are known to be easily and efficiently trainable, even with many deep layers, thanks to residual connections.
Furthermore, ResNets are compatible with fine-tuning, as required by the curriculum learning where the network weights need to be readjusted progressively with increasing the difficulty.

Our ResNet model takes as input the game state $s_t$, which is split into two flatten arrays, one for the real part of $s_t$, one for the imaginary part.
The architecture of the model includes a main torso of $5$ residual blocks, each with $64$ channels for feature extraction and a kernel size of $3 \times 3$. 
The network then split into two distinct heads for policy and value predictions.
The policy head first applies a convolutional layer with $32$ channels and kernel size of $1\times 1$, before processing the output using a dense layer and a softmax activation function to produce a probability distribution over all actions $a\in A_s$.
Similarly, the value head uses a convolutional layer with $8$ channels and kernel size of $1\times 1$, followed by a dense layer and a $\tanh$ activation function to output a scalar value in $[-1,1]$.
To stabilize and accelerate training, we add batch normalization layers after each convolutional operation.

The agent is trained using curriculum learning, on average depth ranging from $5$ to $30$. 
For each target depth $d$, we run $5$ epochs, each consisting of $n_\text{games}=2048$ games or $n_\text{steps}=2048\times d$ steps.
We use a batch size of $64$.
At the end of each epochs, we update the neural network weights using \textsc{Adam} with a learning rate of $10^{-3}$. For stability, we also include an $L2$-regularization term with a $10^{-4}$ weight.
We then compare the updated neural network $\Pi_\text{new}$ to the best-performing neural network $\Pi_\text{best}$ by playing $100$ games.
$\Pi_\text{best}$ is replaced by $\Pi_\text{new}$ if the latter can synthesize $5$ more target unitaries that the former.
After completing the $5$ epochs, we increment $d$ by 1.

We train our agents using two NVIDIA Titan Xp GPUs. This setup results in training time of $\approx 1\mathrm{day}19\mathrm{h}$ for the architecture with $2$ data qubits and $1$ ancilla with a restricted qubit connectivity ($2+1$ architecture), $\approx 4\mathrm{day}7\mathrm{h}$ for $3$ data qubits and $1$ ancilla ($3+1$) qubits and a restricted qubit connectivity, and $\approx 4\mathrm{day}10\mathrm{h}$ for 3 qubits without ancilla and a all-to-all connectivity. These 3 cases are discussed further in the next section. We study the carbon impact of the training process in \cref{app:CO2}.

\section{Results}
\label{sec:results}

\subsection{Architecture and gateset explored}

To test our training pipeline, we train agents to synthesize 2 and 3 qubits unitary, using the $\{\text{H},\text{S},\text{T},\text{CNOT}\}$ gate set and considering an all-to-all connectivity. 
This allows us to compare our trained agents against those of Ref.~\cite{Rietsch2024}.

We then train agents on a clean ancilla architectures using $2+1$ and $3+1$ qubits, respectively.
To showcase the versatility of our approach, we consider a restricted  connectivity and specific gate locations where single qubit gates are applied on the ancilla only, and where CNOT are restricted to act between the ancilla and any data qubit.
Note that this connectivity maintains universality -- swap operations can be implemented using CNOT operations.
We constraint random target circuits to a minimum depth of 5 and a maximum depth 40.

\begin{figure}
    \centering
    \includegraphics[width=0.5\textwidth]{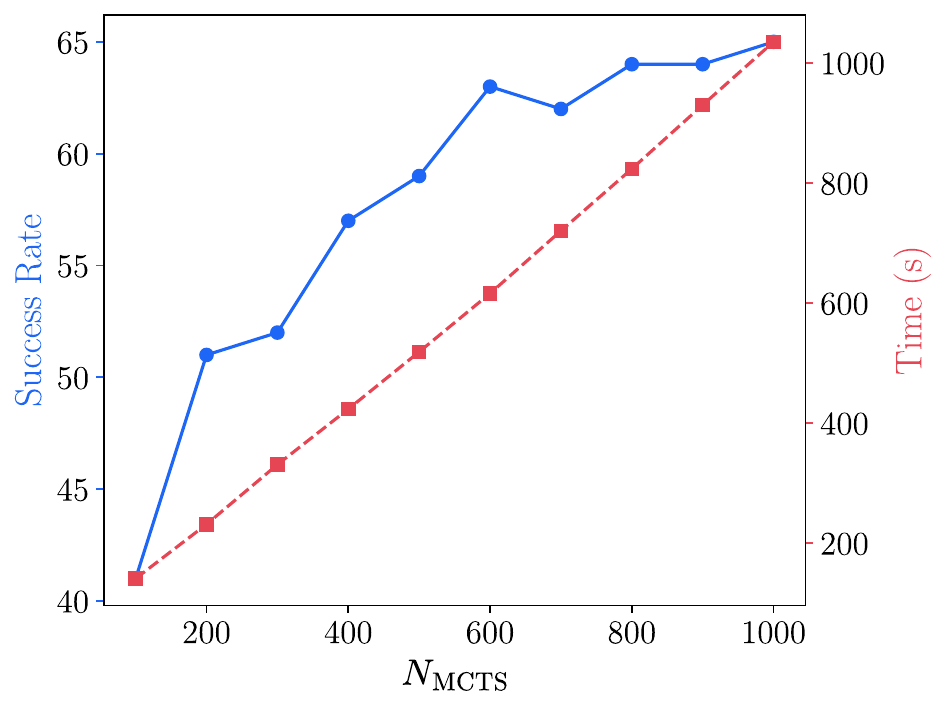}
    \caption{Unitary synthesis success for 100 randomly sampled circuits of depth 10 on a 3+1 architecture, using AlphaZero agents with varying simulation budget, $N_\text{MCTS}$, and a fix temperature $T=0$.}
    \label{fig:n_iters}
\end{figure}

\begin{figure*}
    \centering
    \subfloat[2+1 qubits]{\includegraphics[width=.5\textwidth]{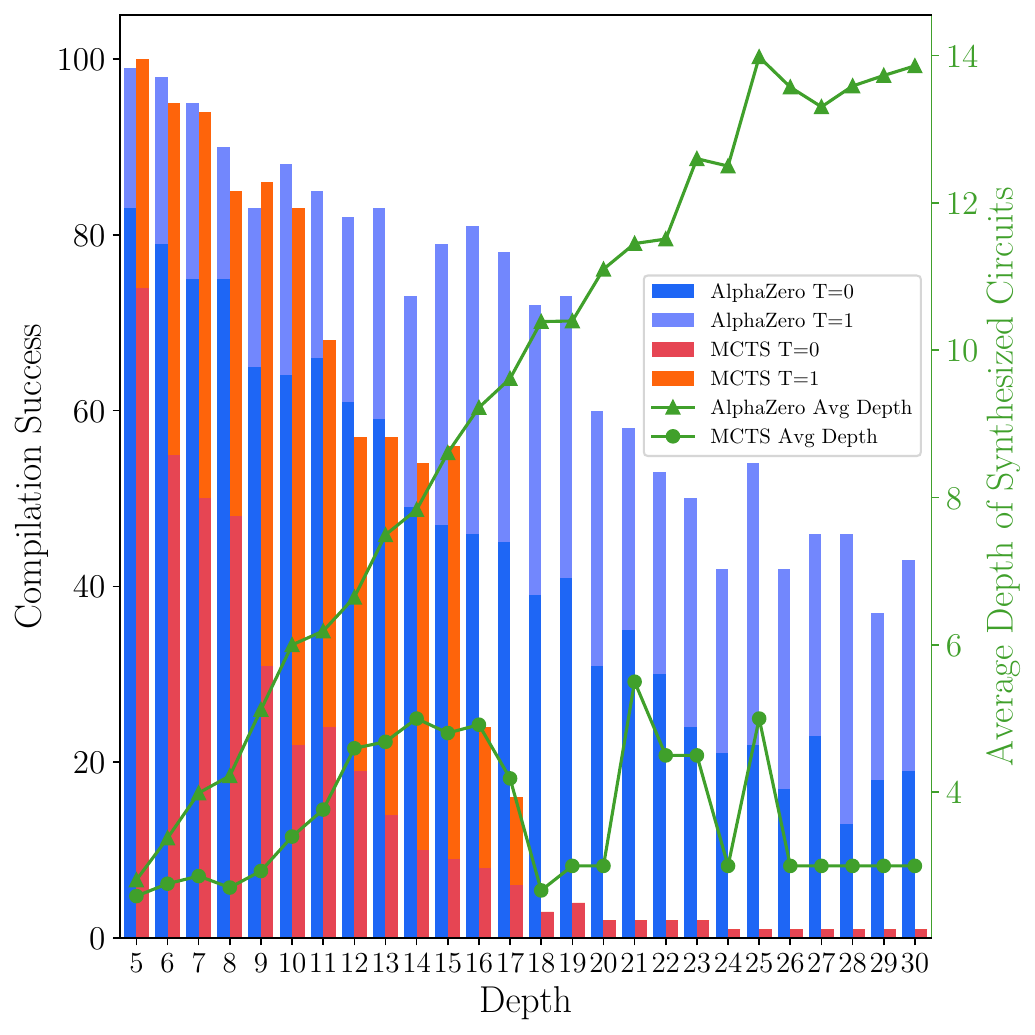}}
    \subfloat[3+1 qubits]{\includegraphics[width=.5\textwidth]{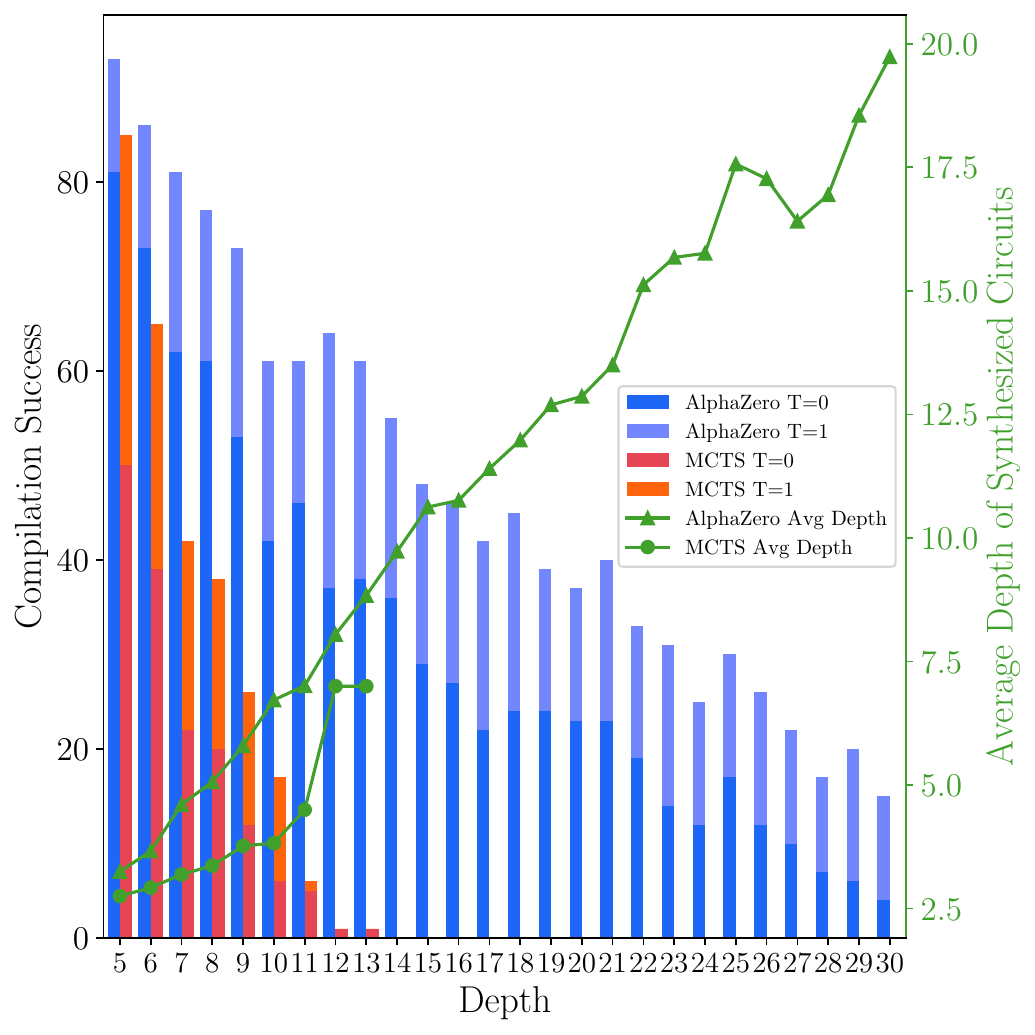}}
    \caption{Performance of trained AlphaZero agents in synthesizing unitaries generated from random circuits of varying depth into circuits of similar or lower depth. For each depth $d$, we attempt to synthesize 100 random target unitaries. An attempt is considered successful if the synthesized circuit is of at most depth $d$. For each target unitary, we first run the AlphaZero agent with $T=0$; if synthesis fails, we give it $10$runs at $T=1$. As a baseline, we run an MCTS with the same two regimes.}
    \label{fig:benchmark}
\end{figure*}

\subsection{Quantitative benchmark}

To evaluate the performance of our trained agents, we propose a benchmark based on randomly sampled target unitaries. 
For each agent, we choose these unitaries from randomly constructed quantum circuits using the same gate set and qubit connectivity the agent has been trained on.
A synthesis is considered successful if it implements the target unitary and if the designed circuit depth is at most the depth of the circuit used to produced the target unitary.
For this benchmark, we set the simulation budget to $N_\text{MCTS}=400$, balancing agent performance with runtime efficiency to enable exploration across a broad range of target unitaries in a reasonnable time.
Note that \cref{fig:n_iters} illustrates the trade-off between agent performance and runtimes for different values of $N_\text{MCTS}$.

For each test unitary, we first attempt synthesis using a deterministic policy ($T=0$). If no circuits were found, we switch to a stochastic policy ($T=1$) and perform 10 independent runs, for an average total runtime of approximately $10s$. Synthesis is considered successful if any of these runs produces a circuit implementing the target unitary.

To benchmark our trained agents we also run a Monte Carlo tree search without neural network guidance.
We keep the simulation budget to be the same.
Similarly, we first run a deterministic MCTS policy with $T=0$, and, if that fails, we run $10$ runs with a temperature $T=1$.

We conduct this benchmark for agents trained for the 2+1 and 3+1 qubits architecture.
The results are shown in \cref{fig:benchmark}.
Our AlphaZero agents outperfomed the MCTS baseline in every case, allowing to synthesize unitaries sampled from high depth circuits.
Comparing the $2+1$ and $3+1$ architectures, we see the benefits of AlphaZero agents against MCTS agents as the number of qubit increases. 
Note that the average depth of the synthesized circuits is always lower than that of the quantum circuit used to produce the target unitaries, hence demonstrating simplification capabilities of our agents.

\subsection{Qualitative benchmark}

As quantitative benchmarks might not be relevant for practical use cases, we compile unitaries that are fundamental elements of quantum operations.
All the unitaries explored here are clearly defined in Ref.~\cite{Williams2011}.
The circuits successfully synthesized by our agents are available in \cref{tab:qcircuit}.

For 2-qubits unitaries, we aim to compile Control-S (CS), Control-T (CT), Control-H (CH), Control-$\sqrt{\text{NOT}}$ (CV), and iSWAP gates.
Except for CT, our agents successfully synthesized all of these gates, both in the all-to-all connectivity and clean ancilla architectures.
These compilations have been done using the deterministic regime, $T=0$, and therefore have a runtime of a few seconds.
The produced circuits matches known optimal decomposition both in term of T-count and circuit depth.

\begin{figure*}
    \centering
    \subfloat[Circuit synthesized by the 3+1 agent. The first 7 gates reduce to the identity operation and the last two $T^\dag$ can be replaced by a $S^\dag$ gate.]{
        \includegraphics[width=\textwidth]{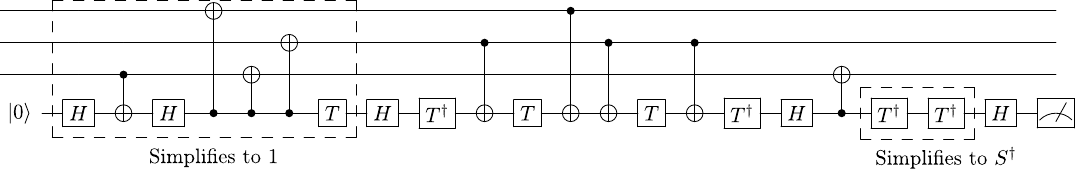}
        \label{fig:toff}
    }
    
    \subfloat[Synthesis of the correction unitary $U_C$ using the 3 qubits all-to-all agent.]{
        \includegraphics[width=\textwidth]{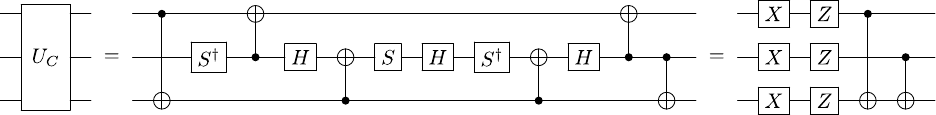}
        \label{fig:correction_toff}
    }
    
    \subfloat[Complete implementation of the Toffoli gate on 3 qubits and 1 clean ancilla, using 4$T$ gates.]{
        \includegraphics[width=\textwidth]{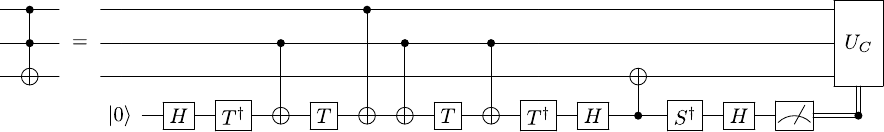}
        \label{fig:final_toff}
    }
    \caption{Using reinforcement learning to synthesize the Toffoli gate for conditionally clean ancilla architecture.}
    \label{fig:toffoli}
\end{figure*}

The CT unitary have been found only in the clean-ancilla regime. This is in accordance with the fact that the CT gate can not be implemented using a Clifford+$T$ gateset on 2 qubits without the use of an ancilla~\cite{Giles2013}. The circuit found matches the one proposed in Ref.~\cite{GidneySO}.

For 3-qubits operations, we attempt to synthesize the Toffoli or CCX, CCZ, and Fredkin or Control-SWAP gates.
In the all-to-all connectivity architecture, we recover known optimal circuits for the Toffoli and CCZ gates.
While synthesizing the CCZ took $3.3s$ using the deterministic $T=0$ regime, finding the Toffoli decomposition took $102.2s$ and took $100$ runs using a temperature $T=0.6$.
Synthesizing the Fredkin operation took 19 gates instead of the known optimal of $17$, and necessitated $1000$ runs at $T=0.6$ for a total runtime of $3042s$.
Using clean-ancilla architecture, we find an implementation of the Toffoli and CCZ gate using $7T$-gates in $200$s runtimes from $200$ runs at $T=0.6$ (see the discussion in the next section for the synthesis of the Toffoli gate). However, our agent failed to synthesized the Fredkin gate in the given $1000$ runs -- or $3000$s of maximum runtime.

\subsection{Improving synthesized circuits}

Trained agents can be used in a human-machine interaction loop, i.e. from a first circuit draft provided by the agent, it is often possible to simplify and improved the circuit.
To illustrate this interaction, we study the compilation of a Toffoli operation using the clean-ancilla architecture. 
In approximately $200s$ runtime, our agent provides the quantum circuit given in \cref{fig:toff}.
From this circuit, we manually discard the first 7 gates, which are equivalent to the identity channel. 
These unnecessary operations are likely caused by the high temperature $T$ used to sample the actions in the first few steps.
We then replace the last two $T^\dag$ gates with a single $S^\dag$ operation.
Note that these two improvements can be algorithmically addressed using identity checks and basic replacement rules.

As the Tofolli implementation proposed in \cref{fig:toff} is conditionned on the ancilla behing measured in $|0\rangle$, we provide the correction operation $U_C$, to apply on the data qubit when the ancilla is measured in the $\ket{1}$ state.
Demonstrating the complementarity of our agents, we use two AlphaZero agents trained on 3 qubits using the $\{\text{H}, \text{S}, \text{T}, \text{CNOT}\}$ and $\{\text{X}, \text{Z}, \text{H}, \text{S}, \text{T}, \text{CNOT}\}$ gatesets, respectively.
In a runtime of less than a second, we find the corrections given in \cref{fig:correction_toff}.
Together, this leads to a deterministic implementation of the Toffoli gate using 4 $T$ gates as depicted in \cref{fig:final_toff}.
This matches the $T$-depth proven optimal in Ref.~\cite{Gouzien2025}, which took a runtime of an hour.

Note that we were not able to retrieve a circuit with a similar T-count using Synthetiq on with a Clifford+T gate and one clean ancilla given a maximum runtime of $200$s. 

\section{Conclusion}

AlphaZero-like agent can be trained to exactly synthesize unitaries with constraints of the physicial computing architecture.
At the cost of a resource intensive training, agents can synthesize target unitaries in short runtimes.
Synthesized circuits can serve as a first basis for furthermore improvements in a human-machine interaction loop or using further automated tooling.
We have shown that agents can learn dynamic circuit and find unusual  implementations of fundamental unitaries.
More broadly, this demonstrates that agents can be trained to exploit new theoretical tools, further advancing the design of quantum circuits.

\acknowledgements
The authors would like to thank Julian Zivy and Elie Gouzien for fruitful discussions. This work is partially supported by the French National program Programme d’investissement d’avenir, IRT Nanoelec, with the reference ANR-10-AIRT-05.


\bibliography{references}


\appendix

\section{Monte Carlo Tree Search as agent} 
\label{app:MCTS}

\subsection{General Monte Carlo Tree Search Algorithm}

The Monte Carlo Tree Search (MCTS) algorithm is a heuristic search algorithm for decision processes~\cite{Coulom2007}. It iteratively explores and constructs a decision tree where each node represents a state, and edges represent actions. By balancing exploration and exploitation, the algorithm selects well-chosen actions to explore valuable regions of the tree.

\medbreak

\begin{figure}[ht]
    \centering
    \includegraphics[width=\linewidth]{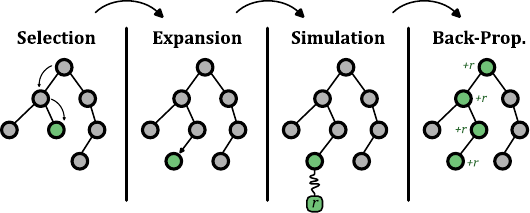}
    \caption{The four steps of Monte Carlo Tree Search.}
    \label{fig:MCTS}
\end{figure}

The MCTS algorithm explores the decision tree through four key steps: selection, expansion, simulation, and back-propagation, as depicted in \cref{fig:MCTS}. Each node in the decision tree represents a state $s$ and stores two pieces of data: the cumulative reward from that state $R_s$, and the number of times the node has been visited $N_s$.
\begin{itemize}
    \item \textbf{Selection Step:}
    starting from the root of the tree, nodes are successively selected until a leaf node (a state that has never been rolled out) is reached. The exploration path is determined by consecutively choosing the branch that maximizes the UCT formula explained below.
    \item \textbf{Expansion Step:}
    this step aims to grow the tree by expanding a new leaf from the leaf node selected in the selection step.
    \item \textbf{Simulation Step:}
    a random game is played from the expanded leaf, resulting in a reward $r$.
    \item \textbf{Back-propagation Step:}
    finally, the statistics of the tree are updated by adding $r$ to the cumulative reward $R_s$ of the visited nodes, and incrementing the visit count $N_s$ by 1.
\end{itemize}

Starting with an empty tree, the algorithm iteratively follows these four steps, expanding the tree one node per iteration. From the back-propagation step, we understand that the data stored by each node is an estimation of how valuable is the node. More formally, for a node $s$, the quantity $\frac{R_s}{N_s}$ can be seen as an expectation of the reward if a random game is played from this node.

\medbreak

The power of MCTS relies on the balance between exploration and exploitation. Generally, the main difficulty in exploring a large decision tree is finding a path that leads to a valuable area of the tree. The MCTS algorithm addresses this issue by evaluating the potential of the branch using the \textit{Upper Confidence Bound 1 applied to Trees} (UCT)~\cite{Kocsis2006}. During the selection step, an exploration path is decided by consecutively choosing the branch that maximizes the quantity
\begin{equation*}
    UCT_i = Q_i + C \cdot \sqrt{\frac{\ln N}{n_i}},
\end{equation*}
where $i$ denotes the studied branch, $Q_i$ is the expected reward of the child node from branch $i$, $n_i$ is the number of times the child node has been visited, $N$ is the number of times the current node has been selected, and $C$ is a balancing constant. This formula illustrates the competition between exploitation and exploration. For instance, a known rewarding area of the tree will have high values of $Q_i$ and $n_i$ -- as it leads to rewarding games and has been visited many times -- resulting in the dominance of the term $Q_i$. Conversely, an unknown area of the tree will have low values of $Q_i$ and $n_i$, resulting in the dominance of the term $\sqrt{\frac{\ln N}{n_i}}$.

\subsection{Monte Carlo Tree Search as an RL agent}

\begin{figure}[ht]
    \centering
    \includegraphics[width=\linewidth]{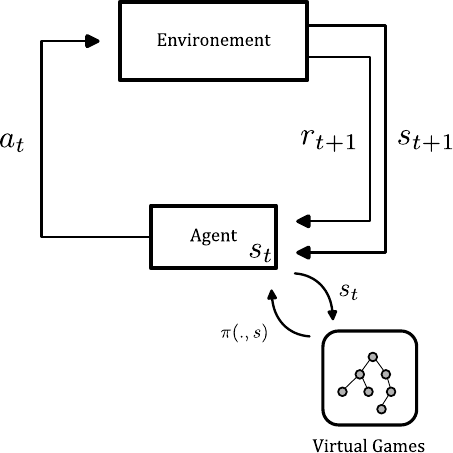}
    \caption{Framework of an Monte Carlo Tree Search agent. Every step $t$ the agent uses a virtual environment to play $N_\text{MCTS}$ virtual games using the MCTS algorithm, resulting a policy $\pi\left(a_t, s_t\right)$ for the action choice $a_t$.}
    \label{fig:RL_MCTS_framework}
\end{figure}

In the context of reinforcement learning, Monte Carlo Tree Search can serve as a policy provider for the learning agent. At each step of the game, the agent simulates games through MCTS iterations, exploring the tree and potential outcomes from the current state. These simulated games are referred to as \textit{virtual games}. To some extent, MCTS can be seen as a \textit{virtual environment} accessible to the agent, where the agent can play some virtual games in order to best predict its next move (c.f. \cref{fig:RL_MCTS_framework}).

\medbreak

More formally, at each step $t$ of the game, the agent plays $N_{MCTS}$ virtual games from the current state $s_t$, constructing a decision tree starting from $s_t$, as the root node, and with at most $N_{MCTS}$ nodes. Through these simulations, the agent gathers statistics and builds a policy
\begin{equation*}
    \pi\left(a, s_t\right) = \frac{n\left(a, s_t\right)}{N\left(s_t\right)},
\end{equation*}
where $N\left(s_t\right)$ is the number of times the node $s_t$ has been visited, and $n\left(a, s_t\right)$ is the number of visits to the child node resulting from action $a$. Furthermore, MCTS provides an approximation of the $Q$-function of the current state $Q_{\pi}\left(., s_t\right)$ for the MCTS policy $\pi$
\begin{equation*}
    Q_{\pi}\left(a, s_t\right) = \frac{R\left(a, s_t\right)}{n\left(a, s_t\right)},
\end{equation*}
where $R\left(a, s_t\right)$ and $n\left(a, s_t\right)$ are the total reward and the number of visits of the child node resulting from action $a$ from $s_t$. From these quantities, the agent can estimate the expected cumulative reward of the current state, known as the value function:
\begin{equation*}
    V_{\pi}\left(s_t\right) = \sum_{a} \pi\left(a, s_t \right) \cdot Q_{\pi}\left(a, s_t\right).
\end{equation*}

\medbreak

This efficient method of building a policy and choosing an action can be viewed as a prediction of the agent's future moves: after constructing a tree of depth $l$, the agent has predictions for the value function of the next $l$ moves. However, while effective, MCTS is a computationally intensive algorithm, as the agent plays $N_{MCTS}$ virtual games at each step. MCTS agents are therefore prohibitive when games are slow to simulate.

\section{Alphazero agent} 
\label{app:Alphazero}

While being effective, AlphaZero proposes an improved version Monte Carlo Tree Search where a deep neural network is used for reward and policy predictions. It addresses the computationally intensive aspect of MCTS while improving its search depth.

\subsection{Hybrid version of the MCTS}

For better efficiency, AlphaZero employs a hybrid version of the MCTS algorithm where a \textit{neural network} is integrated with the tree search for reward and policy predictions. The neural network integration reduces the heavy computational aspect of MCTS while enabling smarter and deeper exploration of the tree. In this context, the neural network acts as a policy provider; for an input state $s$, the neural network outputs a policy $\pi_{s}$ and value function $V_{\pi}$ used in  the \textit{selection step} and the \textit{simulation step} of the MCTS algorithm.

\medbreak

First, during the \textit{selection step}, the neural network is called to the choose the selection path and enhance the tree exploration, allowing smarter exploration. AlphaZero actually uses a hybrid version of the UCT formula where the potential of a branch $a$ from state $s$ is evaluated with the neural network through
\begin{equation*}
    UCT\left(a,s\right) = Q\left(a,s\right) + c_{puct} \pi_s^\prime(a) \cdot \frac{\sqrt{N(s)}}{n\left(a,s\right)+1}
\end{equation*}
where $Q\left(s,a\right)$ is the expected reward of the child node resulting from the branch $a$ of the node $s$, $c_{puct}$ is a balance constant, $\pi_s^\prime\left(a\right)$ is the prior probability obtained from the neural network, $n\left(a, s\right)$ is the number of times the child node has been visited, and $N(s)$ is the number of times the parent node has been visited. To encourage exploration, a Dirichlet noise $\eta\left(\alpha\right)$ is added to the prior probability such that:
\begin{equation*}
    \pi_s^\prime(a) = (1-\epsilon) \cdot \pi_s(a) + \epsilon \cdot \eta(\alpha)
\end{equation*}
where $\pi(s,a)$ is the policy from the neural network, and $\epsilon \in [0, 1]$ controls the balance.

\medbreak

Second, during the \textit{simulation step}, the agent uses the neural network to provide a prediction of the reward instead of playing a game with random moves, significantly reducing computational time in resource-intensive environments. 

\subsection{Learning Protocol}

As opposed to MCTS, AlphaZero agents requires a training phase to produce a neural network with an accurate policy $\pi(\cdot,s)$. Trained on numerous games called \textit{self-played} games, the neural network aims to mimic MCTS behavior by providing an estimation of the policy $\pi$ and value function $V_\pi$. Based on the hybrid version of the MCTS algorithm, the self-play games serve the purpose of decision tree generation through the hybrid MCTS for the neural network's learning. This approach allows the neural network to iteratively refine its predictions while concurrently exploring more valuable regions of the tree. Once trained, the neural network is used for real games as a policy provider in the hybrid version of MCTS.

\medbreak

More formally, during the training phase, the agent plays self-played games using the hybrid MCTS algorithm. Since for every steps the agent plays virtual games via MCTS for decision making, each game produces $n_{steps}$ decision trees. After $n_{games}$, the data generated from the MCTS algorithms are used to update the neural network through back-propagation. A few competition games between the old and updated neural networks assess the improvement, and the most efficient neural network is retained. This process is repeated $n_{epoch}$ times. For better clarity, the \cref{alg:AlphaZero} shows the pseudo-code of this protocol, the function used are :
\begin{itemize}
    \item $MCTS(s, N_\text{MCTS})$: The agent uses the virtual environment to run the MCTS algorithm, with $N_\text{MCTS}$ iterations, from the current state $s$. It outputs an estimation of the policy $\pi$ and value function $V_\pi$ from the statistics.
    \item $AddStorage(\pi, V)$: The policy $\pi$ and value function $V_\pi$ are added to the storage.
    \item $Pick(\pi)$: The agent selects an action according to the policy $\pi$.
    \item $Environment(a, s)$: The agent plays the chosen move by calling the environment.
    \item $UpdateNeuralNetwork()$: Update the parameters of the neural network through back-propagation of the data.
    \item $Competition()$: Competition between the old and updated neural networks; the best one is kept.
    \item $ResetData()$: Reset the data for a new epoch.
\end{itemize}

\begin{algorithm}[H]
\caption{Training phase of AlphaZero}
\label{alg:AlphaZero}
\begin{algorithmic}[1]
\For{$i \gets 1$ to $n_{epoch}$}
    \State $data \gets ResetData()$
    \For{$j \gets 1$ to $n_{game}$}
        \State $s \gets s_{start}$
        \While{$s \ne s_{stop}$}
            \State $\pi, V \gets MCTS(s, n_{MCTS})$
            \State $data \gets AddStorage(\pi, V)$
            \State $a \gets Pick(\pi)$
            \State $s \gets Environment(a, s)$
        \EndWhile
    \EndFor
    \State $UpdateNeuralNetwork(data)$
    \State $Competition()$
\EndFor
\end{algorithmic}
\end{algorithm}

\begin{figure}[ht]
    \centering
    \includegraphics[width=\linewidth]{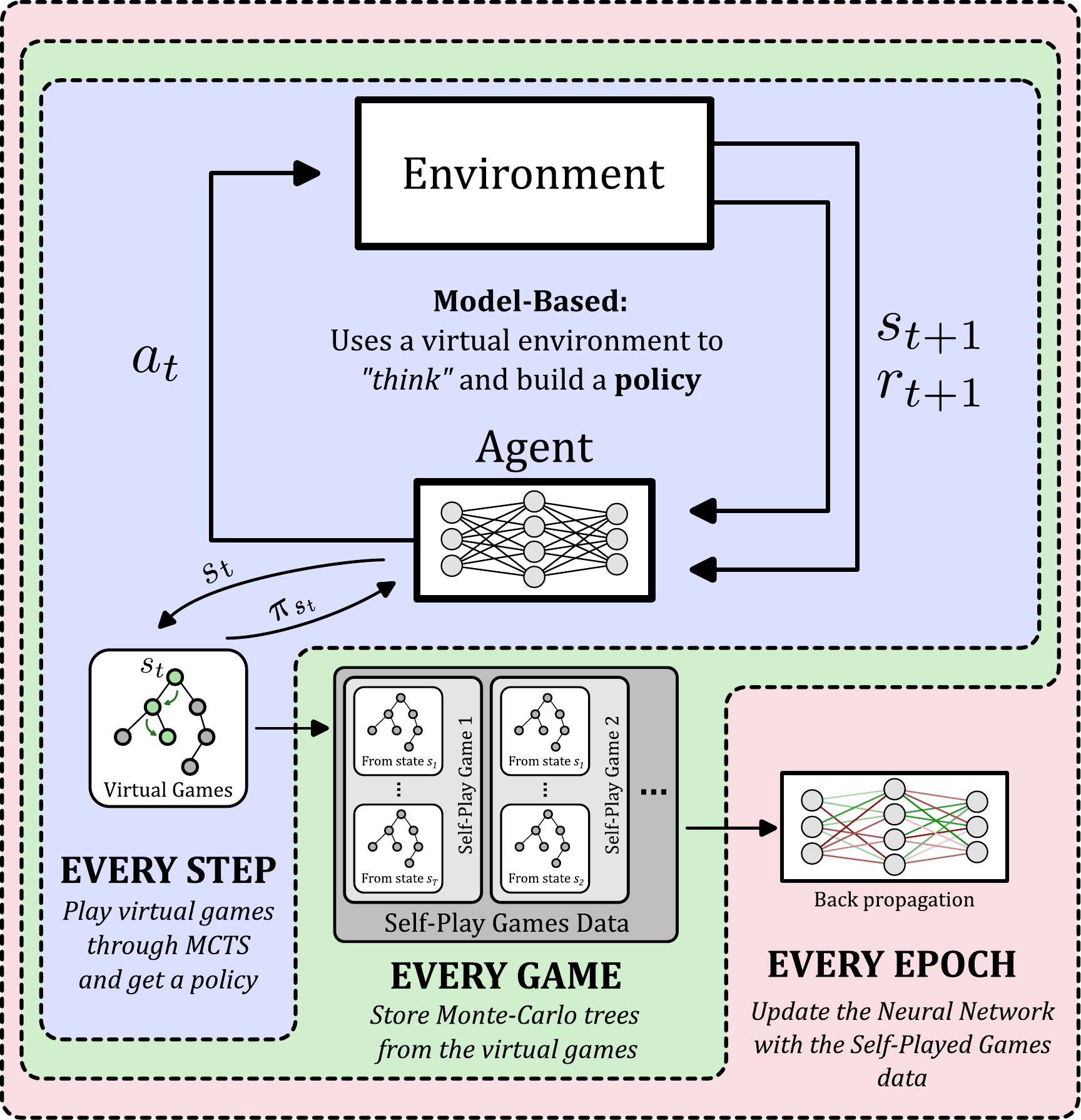}
    \caption{Framework of the training phase of AlphaZero.}
    \label{fig:AZ_framework}
\end{figure}

\section{Sampling random circuit}
\label{app:random_circuit}

Our AlphaZero agents requires target unitaries for the self-play learning phase.
We build these unitaries from randomly sampled quantum circuits using the same gateset $\mathcal{G}$ available to the agent, ensuring a decomposition is possible.

First, to prevent redundant or trivial gate occurring in sampled circuits, we build a \textit{redundancy} list.
For every gate $g_i \in \mathcal{G}$, a boolean list contains all gates $g_j\in \mathcal{G}$ that satisfy the relation
\begin{equation}
    g_i g_j = \id.
\end{equation}
Similarly, we compute a \textit{commutation} list, a boolean list which, for every gate $g_i\in G$, contains every gate $g_j \in G$ satisfying the relation
\begin{equation}
    g_i g_j - g_j g_i = 0.
\end{equation}
Note that these lists are computed only once and stored, in order to save computational resources.

To sample a quantum circuit, we first fix the depth $d$, i.e. the number of gates.
We then randomly pick a gate $g_1$ from $\mathcal{G}$ using a uniform distribution and append it to the quantum circuit.
For each of the remaining $k-1$ gates, we randomly pick a gate $g_k$ and perform a redundancy check. 
We commute $g_k$ with every gate in the quantum circuit in reverse order, using the commutation list. 
If $g_k$ commute with a particular gate, we check whether that gate is in the redundancy list of $g_k$. If it is, we sample another gate $g_k$ until no redundancies are detected.

In the case of clean-ancilla circuits, we sample a random circuit using the same method on the $n+1$ qubit. Once a quantum circuit is sampled, we perform an additional check. Indeed, we need to ensure that the sampled circuit, with the ancilla initialized and measured in $\ket{0}$, produces a unitary. This might not occur, for example, if the quantum circuit sets the ancilla to $\ket{1}$.
Let $U\in \mathcal{C}^{2n \times 2n}$ the unitary produced by the sampled circuit with the ancilla prepared and measured in $\ket{0}$, we verify that 
\begin{equation}
    \frac{UU^\dagger}{\Tr[UU^\dagger]}=\id_n
\end{equation}
If this condition is not satisfied, we sample a new quantum circuit until a unitary in produced.

\section{Carbon emission}
\label{app:CO2}

To train our agents, we use two NIVIDA Titan XP GPUs. Our implementation makes uses of the maximum capacity of the GPUs, resulting in a consumption of $\approx 250W$ per GPU.
In France, for the year 2024, the average CO2 equivalent per KWh is $10g$~\cite{EDF2024}.
The longest training process took $\approx 106\mathrm{h}$, which outputs a CO2 equivalent of $2\times 0.250 \times 10 \times 106 = 0.53kg CO2_{eq}$.

\clearpage

\begin{turnpage}
\begin{table}
\begin{tabular}{|c|c|c|c|}
\hline
\textbf{Qubits} & \textbf{Unitary} & \textbf{$n$+1}, $\{H,T,CNOT\}$ & \textbf{all-to-all}, $\{H,S,T,CNOT\}$ \\ \hline
\multirow{15}{*}{2} & \multirow{3}{*}{CS} &
\Qcircuit @C=0.5em @R=0.5em {
    & \qw & \qw & \qw & \qw & \ctrl{2} & \qw & \qw & \qw & \ctrl{2} & \qw  & \qw \\
    & \qw & \qw & \qw & \qw & \qw & \qw & \ctrl{1} & \qw & \qw & \qw & \qw\\
    & & \lstick{\ket{0}} & \gate{H} & \gate{T} & \targ & \gate{T^\dag} & \targ & \gate{T} & \targ & \meter &
} &
\Qcircuit @C=0.5em @R=0.5em {
     & \gate{T} & \ctrl{1} & \qw & \ctrl{1} & \qw & \qw \\
     & \qw & \targ & \gate{T^\dag} & \targ & \gate{T} & \qw \\
} \\ \cline{2-4}

& \multirow{3}{*}{CT} &
\Qcircuit @C=0.5em @R=0.5em {
    & \qw & \qw & \qw & \qw & \ctrl{2} & \qw & \qw & \qw & \ctrl{2} & \qw & \qw & \qw & \qw & \qw \\
    & \qw & \qw & \qw & \qw & \qw & \qw & \ctrl{1} & \qw & \qw & \qw & \qw & \qw & \qw & \qw \\
    & & \lstick{\ket{0}} & \gate{H} & \gate{T} & \targ & \gate{T^\dag} & \targ & \gate{T} & \targ & \gate{T^\dag} & \gate{H} & \gate{T^\dag} & \gate{H} & \meter
} & \multirow{3}{*}{No circuit found} \\ \cline{2-4}

& \multirow{3}{*}{CH} &
\Qcircuit @C=0.5em @R=0.5em {
    & \qw & \qw & \qw & \ctrl{2} & \qw & \ctrl{2} & \qw & \qw & \qw & \ctrl{2} & \qw & \qw & \ctrl{2} & \qw & \qw & \qw & \ctrl{2} & \qw & \qw & \ctrl{2} & \qw & \qw \\
    & \qw & \qw & \qw & \qw & \qw & \qw & \qw & \ctrl{1} & \qw & \qw & \qw & \qw & \qw & \qw & \targ & \qw & \qw & \qw & \qw & \qw & \qw & \qw \\
    & & \lstick{\ket{0}} & \gate{H} & \targ & \gate{T} & \targ & \gate{T^\dag} & \targ & \gate{H} & \targ & \gate{T} & \gate{H} & \targ & \gate{H} & \ctrl{-1} & \gate{T} & \targ & \gate{H} & \gate{\textcolor{red}{S}} & \targ & \qw & \meter
} & 
\Qcircuit @C=0.5em @R=0.5em {
    & \qw & \qw & \qw & \qw & \ctrl{1} & \qw & \qw & \qw & \qw \\
    & \qw & \gate{S} & \gate{H} & \gate{T} & \targ & \gate{T^\dag} & \gate{H} & \gate{S^\dag} & \qw
}
\\ \cline{2-4}

& \multirow{3}{*}{CV} &
\Qcircuit @C=0.5em @R=0.5em {
& \qw & \qw & \ctrl{2} & \qw & \qw & \ctrl{2} & \qw & \ctrl{2} & \qw & \qw & \ctrl{2} & \qw & \qw & \qw & \qw \\
& \qw & \qw & \qw & \qw & \qw & \qw & \qw & \qw & \qw & \qw & \qw & \targ & \qw & \qw & \qw \\
& & \lstick{\ket{0}} & \targ & \gate{H} & \gate{T^\dag} & \targ & \gate{T^\dag} & \targ & \gate{ {\textcolor{red}{S} }} & \gate{H} & \targ & \ctrl{-1} & \gate{H} & \qw & \meter \\
} & 
\Qcircuit @C=0.5em @R=0.5em {
	& \qw & \gate{T} & \ctrl{1} & \qw & \ctrl{1} & \qw \\
	& \gate{H} & \gate{T} & \targ  & \gate{T^\dag} & \targ & \gate{H}
}
\\ \cline{2-4}

& \multirow{3}{*}{iSWAP} &
\Qcircuit @C=0.5em @R=0.5em  {
    & \qw & \qw & \qw & \ctrl{2} & \qw & \qw & \targ & \qw & \qw & \qw \\
    & \qw & \qw & \ctrl{1} & \qw & \targ & \qw & \qw & \qw & \qw & \qw \\
    & & \lstick{\ket{0}} & \targ & \targ & \ctrl{-1} & \gate{\textcolor{red}{S}} & \ctrl{-2} & \gate{H} & \qw & \meter
} & 
\Qcircuit @C=0.5em @R=0.5em{
	& \qw & \targ & \gate{S} & \ctrl{1} & \targ & \qw \\
	& \qw & \ctrl{-1} & \qw & \targ & \ctrl{-1} & \qw
} \\ \hline

\multirow{10}{*}{3} & \multirow{4}{*}{Toffoli} &
\multirow{4}{*}{See \cref{fig:toffoli}} & 
\Qcircuit @C=.5em @R=.5em  {
& \gate{T^\dag} & \ctrl{2} & \qw & \ctrl{2} & \qw & \qw & \ctrl{1} & \qw & \qw & \qw & \ctrl{1} & \qw \\
& \qw & \qw & \gate{T^\dag} & \qw & \targ & \gate{T} & \targ & \gate{T^\dag} & \targ & \gate{T} & \targ & \qw \\
& \gate{H} & \targ & \gate{T} & \targ & \ctrl{-1} & \qw & \qw & \qw & \ctrl{-1} & \gate{T^\dag} & \gate{H} & \qw
}
\\ \cline{2-4}

& \multirow{4}{*}{CCZ} &
\Qcircuit @C=.5em @R=.5em  {
& \qw & \qw          & \ctrl{3} & \qw & \qw & \qw & \ctrl{3} & \qw & \ctrl{3} & \qw & \ctrl{3} & \qw & \ctrl{3} & \qw & \ctrl{3} & \qw & \qw & \qw & \ctrl{3} & \qw & \qw & \qw \\
& \qw & \qw          & \qw & \qw & \qw & \qw & \qw & \qw & \qw & \qw & \qw & \ctrl{2} & \qw & \qw & \qw & \qw & \qw & \qw & \qw & \qw & \ctrl{2} & \qw \\
& \qw & \qw          & \qw & \qw & \qw & \ctrl{1} & \qw & \qw & \qw & \qw & \qw & \qw & \qw & \qw & \qw & \qw & \ctrl{1} & \qw & \qw & \qw & \qw & \qw \\
& & \lstick{\ket{0}} & \targ & \gate{T^\dag} & \gate{H} & \targ & \targ & \gate{T} & \targ & \gate{T^\dag} & \targ & \targ & \targ & \gate{T} & \targ & \gate{T^\dag} & \targ & \gate{T} & \targ & \gate{T^\dag} & \targ & \meter
} &
\Qcircuit @C=.5em @R=.5em  {
& \gate{T^\dag} & \targ & \gate{T} & \targ & \qw & \qw & \ctrl{1} & \qw & \qw & \qw & \ctrl{1} & \qw \\
& \qw & \qw & \gate{T^\dag} & \qw &\targ & \gate{T} & \targ & \gate{T^\dag} & \targ & \gate{T} & \targ & \qw \\
& \gate{T^\dag} & \ctrl{-2} & \qw & \ctrl{-2} & \ctrl{-1} & \qw & \qw & \qw & \ctrl{-1} & \qw & \qw & \qw
}
\\ \cline{2-4}

& \multirow{4}{*}{Fredkin} &
\multirow{4}{*}{No circuit found}
&
\Qcircuit @C=.5em @R=.5em  {
& \qw & \qw & \qw & \gate{T^\dag} &\qw & \targ & \gate{T^\dag} & \ctrl{1} & \qw & \gate{S} & \targ & \qw &\ctrl{1} & \qw & \qw \\
& \targ & \gate{T^\dag} & \targ & \gate{T} & \targ & \qw & \qw & \targ & \gate{T^\dag} & \targ & \qw & \gate{T} & \targ & \targ & \qw \\
& \ctrl{-1} & \gate{H} & \ctrl{-1} & \gate{T^\dag} & \ctrl{-1} & \ctrl{-2} & \qw & \qw & \qw & \ctrl{-1} & \ctrl{-2} & \qw & \gate{H} & \ctrl{-1} & \qw
} \\ \hline

\end{tabular}
\caption{Synthesized quantum circuits for some fundamental $n=2$ and $n=3$ qubits unitaries. The $n+1$ column corresponds to clean-ancilla architecture circuits build from a $\{H,T,CNOT\}$ gateset. We manually replaced two consecutive $T$-gates by a $S$ gate, labeled in red. The $\textbf{all-to-all}$ column contains circuits build from a $\{H,S,T,CNOT\}$ gateset and using an all-to-all connectivity. } 
\label{tab:qcircuit}
\end{table}
\end{turnpage}

\end{document}